\documentclass[aps,prl,amsmath,amssymb,floatfix,twocolumn,amsmath,superscriptaddress,twocolumn,nofootinbib,tighten,letterpaper]{revtex4}
\usepackage{multirow}
\usepackage{bbold}
\usepackage{subfigure}
\usepackage{color}
\usepackage{mathrsfs}
\usepackage{hyperref}
\usepackage[normalem]{ulem}
\usepackage{bm}

\usepackage{amsfonts, relsize, color}
\usepackage{graphics}
\usepackage{graphicx}
\usepackage{subfigure}
\usepackage{hyperref}
\usepackage{color}
\usepackage{comment}

\begin{document}

\title{Out of equilibrium higher-order topological insulator: Floquet engineering and quench dynamics}

\author{Tanay Nag}~\email{tnag@sissa.it}
\affiliation{SISSA, via Bonomea 265, 34136 Trieste, Italy}
\affiliation{Max-Planck-Institut f\"{u}r Physik komplexer Systeme, N\"{o}thnitzer Str. 38, 01187 Dresden, Germany}

\author{Vladimir Juri\v ci\' c}~\email{vladimir.juricic@nordita.org}
\affiliation{Nordita, KTH Royal Institute of Technology and Stockholm University, Roslagstullsbacken 23,  10691 Stockholm,  Sweden}

\author{Bitan Roy}~\email{bir218@lehigh.edu}
\affiliation{Max-Planck-Institut f\"{u}r Physik komplexer Systeme, N\"{o}thnitzer Str. 38, 01187 Dresden, Germany}
\affiliation{Department of Physics, Lehigh University, Bethlehem, Pennsylvania, 18015, USA}

\date{\today}
\begin{abstract}
Higher-order topological~(HOT) states,~hosting topologically protected modes on lower-dimensional boundaries,~such as hinges and corners,  have recently extended the realm of the static topological phases.~Here we demonstrate the possibility of realizing a two-dimensional \emph{Floquet} second-order topological insulator, featuring corner-localized zero quasienergy modes and characterized by quantized Floquet qudrupolar moment $Q^{\rm Flq}_{xy}=0.5$, by periodically kicking a quantum spin Hall insulator (QSHI) with a discrete fourfold ($C_4$) and time-reversal (${\mathcal T}$) symmetry breaking Dirac mass perturbation.~Furthermore, we show that $Q^{\rm Flq}_{xy}$ becomes independent of the choice of origin as the system approaches the thermodynamic limit.~We also analyze the dynamics of a corner mode after a sudden quench, when the $C_4$ and ${\mathcal T}$ symmetry breaking perturbation is switched off, and find that the corresponding survival probability displays periodic appearances of complete, partial and no revival for long time, encoding the signature of corner modes in a QSHI.~Our protocol is sufficiently general to explore the territory of dynamical HOT phases in insulators and gapless systems.
\end{abstract}

\maketitle

\emph{Introduction}. Topological states of matter in equilibrium are characterized by the bulk-boundary correspondence: A nontrivial topological phase in two (three) dimensions supports gapless edge (surface) modes of codimension \emph{one}, protected by the bulk topological invariant~\cite{Hasan-Kane-RMP2010,Qi-Zhang-RMP2011}. This principle is operative for gapped, such as insulators and superconductors, as well as to the gapless topological states, which for instance include Dirac, Weyl and nodal loop semimetals~\cite{armitage-RMP2018}. The realm of topological states also encompasses the systems out of equilibrium, with Floquet topological insulators realized in the periodically driven systems standing as its paradigmatic representative~\cite{galitski, gedik, moessner, berg, eckardt, takashi, alu, azameit}. In such systems, the bulk-boundary correspondence is more subtle, since bands featuring a trivial static topological invariant may sustain topological boundary modes, due to the nontrivial winding of the wavefunctions in the time direction.

Recently, the notion of topological states in static systems was extended to so-called higher-order topological (HOT) phases, featuring gapless modes on the boundaries of codimension ($d_c$) higher than one~\cite{benalcazar2017, schindler2018, song2017, benalcazar-prb2017, langbehn2017, schindler-sciadv2018, ezawa2018, hsu2018, trifunovic2019, wang1-2018, yan2018, calugaru2019, ahn2018, Klinovaja2019, szabo-HOT}. For example, a three-dimensional second- (third-) order topological insulator hosts gapless modes on the hinges (at the corners), characterized by $d_c=2\;(3)$, in contrast to its conventional or first-order counterpart that accommodates two-dimensional massless Dirac fermions on the surface with $d_c=1$. While elemental Bi has emerged as the prominent candidate for a three-dimensional second order topological insulator~\cite{schindler2018}, HOT states can also be realized in other, noncrystalline, setups~\cite{serra-garcia2018, noh2018, peterson, imhof2018, agarwala, zhang2019, kempkes2019}. A set of fundamentally important questions then arises quite naturally: (a) Can Floquet HOT phases be engineered via periodic driving? (b) Can HOT phase leave any signature after a sudden quench to a lower order phase? Due to experimental advancements on the Floquet techniques~\cite{gedik, alu, azameit,Exp-1,Exp-2} and quench dynamics~\cite{Exp-3, Exp-4}, extending the reach of HOT phases to dynamical (out-of-equilibrium) systems has also become experimentally pertinent. This is the quest we seek to venture in this work, which is still at its infancy~\cite{Wu2018,Gong2019,Seradjeh2018,diptiman2019, Gilrefael2019}.

\begin{figure}[t!]
\includegraphics[width=\linewidth]{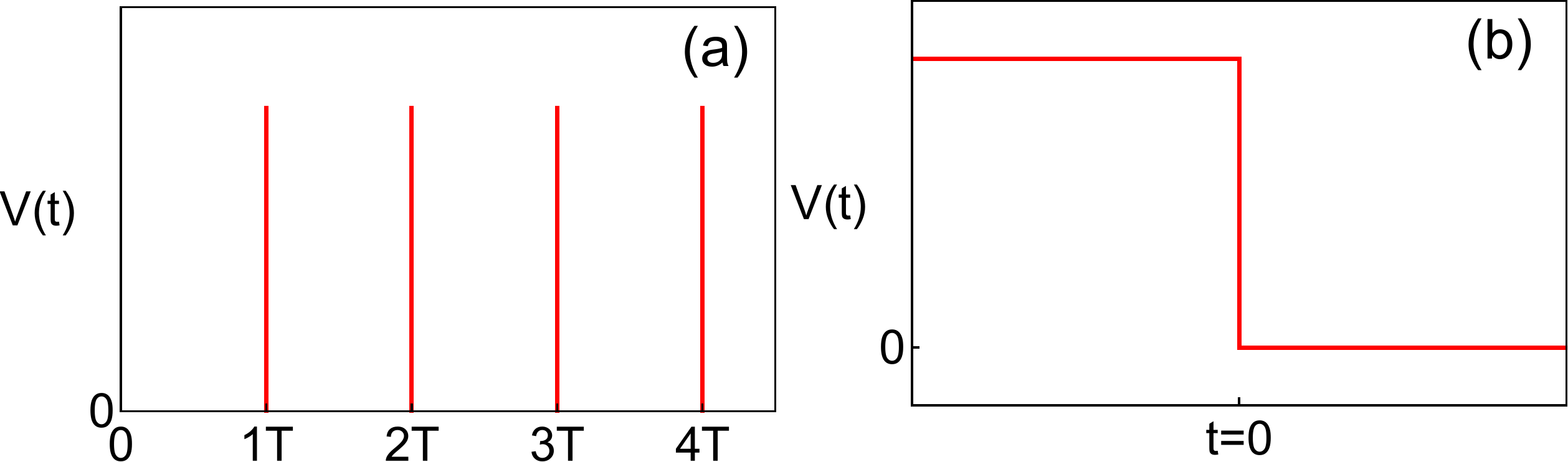}
\caption{(a) Schematic representation of a periodic kick by $C_4$ and ${\mathcal T}$ symmetry breaking perturbation $V(t)$ in a quantum spin Hall insulator, given by Eq.~(\ref{Eq:periodickick}), supporting one-dimensional edge modes [see Fig.~\ref{Fig:FloquetHOTI_Summary}(b)], giving rise to corner-localized quasimodes at stroboscopic time $t=T$ [see Fig.~\ref{Fig:FloquetHOTI_Summary}(d)]. (b) Sudden quench of such a perturbation at time $t=0$, as described by Eq.~(\ref{Eq:Profile_Suddenquench}), leading to the survival probability of one corner mode for $t>0$ (see Figs.~\ref{Fig:survivalprob_summary} and \ref{Fig:wavefunction_quench}).}
\label{Fig:FloquetKickQuench}
\end{figure}

\begin{figure*}[t!]
\includegraphics[width=\linewidth]{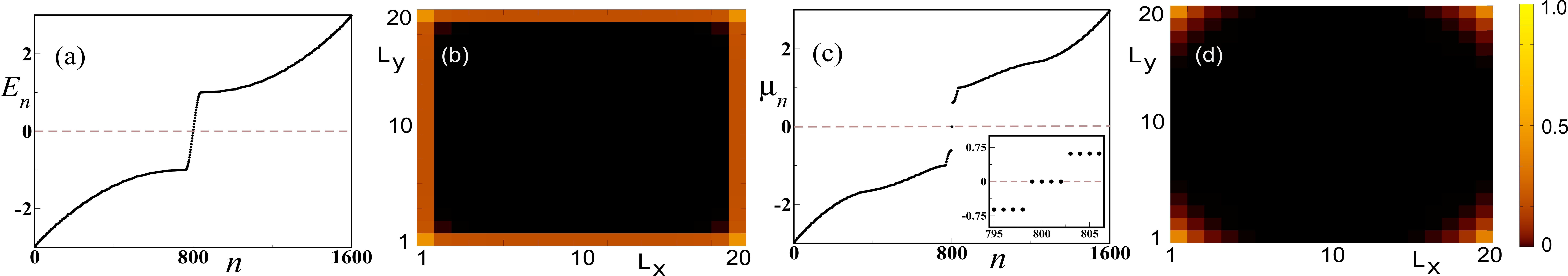}
\caption{Floquet engineering of a two-dimensional second-order topological insulator, supporting four zero quasienergy ($\mu_n=0$) corner modes [see (c) and inset], from an initial QSHI that hosts one-dimensional chiral edge modes of energy $E_n=0$ [see (a)]. In (a) and (c), $n$ is the index for energy ($E_n$) and quasienergy ($\mu_n$), respectively. Local density of states: (b) the chiral edge mode associated with a static QSHI [see Eq.~(\ref{Eq:QSHI_Hamiltonian})] of energy $E_n=0$, and (d) zero quasienergy states ($\mu_n=0$) of the Floquet operator $U(T)$ [see Eq.~(\ref{Eq:Floquetoperator})] at the stroboscopic time $t=T$, obtained by periodically kicking a QSHI by a $C_4$ and ${\mathcal T}$ symmetry breaking perturbation [see Fig.~\ref{Fig:FloquetKickQuench}(a) and Eq.~(\ref{Eq:periodickick})]. We here set $t_0=t_1=m=1$, $\Delta=0.3$ and $T=0.628$. The corresponding frequency $\omega=2 \pi /T \approx 10 \gg t_{0,1}$, yielding the high frequency regime. 
}~\label{Fig:FloquetHOTI_Summary}
\end{figure*}

\begin{figure}[t!]
\includegraphics[width=1\linewidth,height=.9\linewidth]{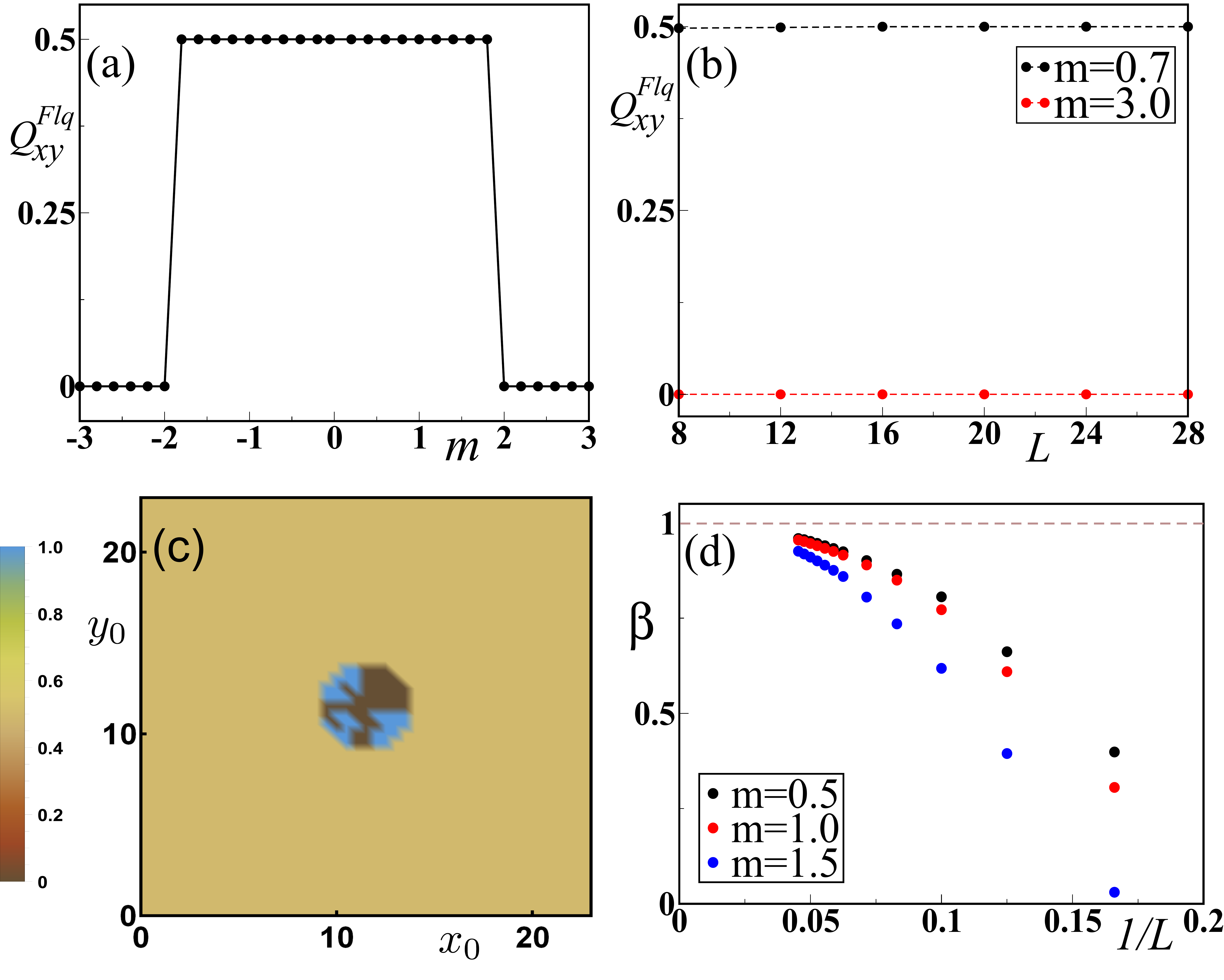}
\caption{(a) Floquet quadrupolar moment ($Q^{\rm Flq}_{xy}$) in a system with linear dimension $L=20$ in the $x$ and $y$ directions, with a choice of origin $(x_0,y_0)=(0,0)$ for $|m|\leq 3$. Floquet HOT (trivial) insulator is realized for $|m|<2 (|m|>2)$ with $Q^{\rm Flq}_{xy}=0.5 (0.0)$. (b) For $(x_0,y_0)=(0,0)$, $Q^{\rm Flq}_{xy}$ is independent of $L$ inside the Floquet HOT ($m=1.0$) and trivial ($m=3.0$) insulating phases. (c) Origin dependence of $Q^{\rm Flq}_{xy}$ in a system with $L=24$ inside a Floquet HOTI (supporting corner modes in open system) for $m=1$. Unless the origin of the system is chosen around its center, we find $Q^{\rm Flq}_{xy}=0.5$. (d) Scaling of the fraction of the area ($\beta$) in the $(x_0,y_0)$ plane, yielding $Q^{\rm Flq}_{xy}=0.5$ for various choices of $m$ that support Floquet HOTI (hosting zero quasienergy corner modes in open system), with $1/L$. As the system approaches the thermodynamic limit ($L \to \infty$), $\beta \to 1$ and $Q^{\rm Flq}_{xy}=0.5$ becomes independent of the choice of the origin. Inside the Floquet trivial phase $Q^{\rm Flq}_{xy}=0.0$ is independent of the choice of origin for any $L$. We always compute $Q^{\rm Flq}_{xy}$ in a periodic system. Similar behavior of the quadrupolar moment ($Q_{xy}$) has already been reported for static crystalline and amorphous HOTI~\cite{agarwala}.
}~\label{Fig:FloquetQuadrupolar}
\end{figure}

\begin{figure}[t!]
\includegraphics[width=0.98\linewidth]{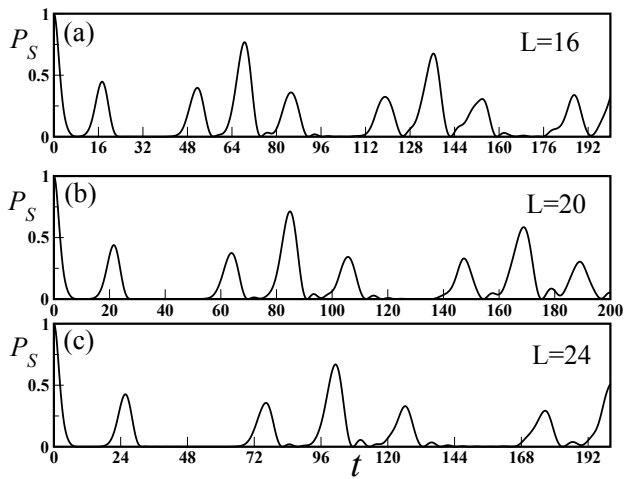}
\caption{Dynamics of the survival probability ($P_s$), see Eq.~(\ref{Eq:SurvivalProb_Def}), of a zero energy corner mode with time ($t$) after a sudden quench at $t=0$, when the $C_4$ and ${\mathcal T}$ symmetry breaking perturbation is completely switched off, see Fig.~\ref{Fig:FloquetKickQuench}(b) and Eq.~(\ref{Eq:Profile_Suddenquench}). Here, $L$ is the linear dimension of the system along both $x$ and $y$ directions. Large (intermediate) peaks correspond to complete (partial) revival, while the (almost) flat regions refer to no revival. For numerical simulation we set $t_0=t_1=m=1$ and $\Delta=0.3$. The spectral density of the corresponding time evolved wavefunction is shown in Fig.~\ref{Fig:wavefunction_quench}.
}~\label{Fig:survivalprob_summary}
\end{figure}

\begin{figure*}[t!]
\includegraphics[width=\linewidth]{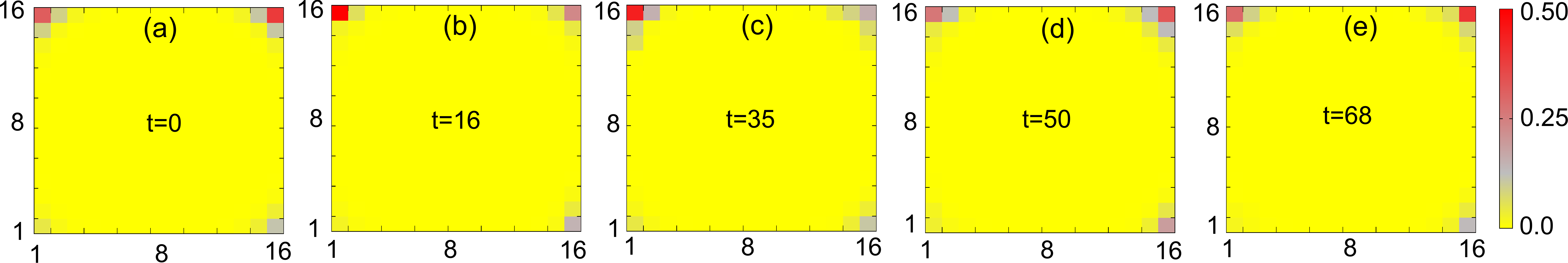}
\caption{ Spectral density of the time evolved state $|\Psi(t) \rangle = \exp(-i H_{\rm Fin} t) |\Psi^{\rm initial}_{\rm corner} \rangle$ at time $t=0$ and various instances after a sudden quench at $t=0$ [see Fig.~\ref{Fig:FloquetKickQuench}(b)] in a system with linear dimension $L=16$ in both directions. Here, $|\Psi^{\rm initial}_{\rm corner} \rangle$ represents a corner mode of the initial Hamiltonian $H_{\rm Ini}=H_{\rm SHI} + \Gamma_4 V_{12}$. Since, the final Hamiltonain $H_{\rm Fin}=H_{\rm SHI}$ accommodates one-dimensional chiral edge modes, propagation of the corner state at time $t>0$ predominantly takes place through the boundaries of the system, leading to the observed dynamics of the survival probability, see Fig.~\ref{Fig:survivalprob_summary}.
}~\label{Fig:wavefunction_quench}
\end{figure*}

In this paper, we promote a general mechanism of engineering Floquet HOT phases by periodically kicking [Fig.~\ref{Fig:FloquetKickQuench}(a)] a first-order topological phase with a suitable symmetry breaking mass perturbation. As a demonstrative example, we show that a two-dimensional second-order Floquet topological insulator, supporting zero quasienergy corner states [Fig.~\ref{Fig:FloquetHOTI_Summary}(d)], can be generated by periodically kicking a quantum spin Hall insulator (QSHI), hosting one-dimensional edge modes [Fig.~\ref{Fig:FloquetHOTI_Summary}(b)], with a $C_4$ and ${\mathcal T}$ symmetry breaking \emph{mass} perturbation. Our proposed general protocol for engineering Floquet HOT phases is distinct from the \emph{three-step} driving procedure in a QSHI~\cite{Wu2018} or a $C_4$ and ${\mathcal T}$ symmetry breaking trivial insulator~\cite{diptiman2019}. The resulting higher order topological insulator (HOTI) is characterized by quantized Floquet quadrupolar moment $Q^{\rm Flq}_{xy}=0.5$ (within numerical accuracy, see Fig.~\ref{Fig:FloquetQuadrupolar}(a)), which is independent of system size ($L$), see Fig.~\ref{Fig:FloquetQuadrupolar}(b). Furthermore, a minor origin dependence of $Q^{\rm Flq}_{xy}$ inside the topological phase [Fig.~\ref{Fig:FloquetQuadrupolar}(c)] disappears as the system approaches the thermodynamic limit ($L \to \infty$), as shown in Fig.~\ref{Fig:FloquetQuadrupolar}(d). We also study the dynamics of a zero-energy corner mode following a quench~\cite{Diptiman_book,sengupta_review,aditimitra_review,calabrese,Adutta,sacramento,tanay}, such that the final state is first order. In particular, we compute its \emph{survival probability} for time $t>0$ [Fig.~\ref{Fig:survivalprob_summary}], after suddenly switching off the $C_4$ and ${\mathcal T}$ symmetry breaking perturbation at $t=0$ [Fig.~\ref{Fig:FloquetKickQuench}(b)]. Due to the edge propagation of the corner mode (see Fig.~\ref{Fig:wavefunction_quench}) in the postquench QSHI phase, the survival probability displays periodic appearances of complete, partial and no revival for long time. Therefore, our results should open up a route to study the Floquet HOT phases and quenching dynamics of HOT states in different setups (such as semimetals) and dimensions.

\emph{Floquet HOTI}. We begin with a two-dimensional QSHI, described by the Hamiltonian
\begin{equation}~\label{Eq:QSHI_Hamiltonian}
H_{\rm SHI}= t_1 \sum^{2}_{j=1} \Gamma_j S_j - t_0  \Gamma_3 \big[m- \sum^2_{j=1}C_j \big]
\equiv {\bf N} ({\bf k}) \cdot {\boldsymbol \Gamma},
\end{equation}
where $S_j \equiv \sin (k_j a)$ and $C_j \equiv \cos(k_j a)$, $a(=1)$ is the lattice spacing, $k_j$s are components of momentum. The four-component mutually anti-commuting Hermitian $\Gamma$ matrices are $\Gamma_1=\sigma_3 \tau_1$, $\Gamma_2=\sigma_0 \tau_2$, $\Gamma_3=\sigma_0 \tau_3$. The Pauli matrices ${\boldsymbol \tau}$ (${\boldsymbol \sigma}$) operate on the sublattice/orbital (spin) degrees of freedom. For $0 < |m|< 2$, the system describes a QSHI. It supports two counterpropagating edge modes for opposite spin projections [see Fig.~\ref{Fig:FloquetHOTI_Summary}(b)], effectively described by two copies of one-dimensional massless Dirac fermions~\cite{BHZ}. We drive such a QSHI by a periodic kick of amplitude $\Delta$ and periodicity $T$ [see Fig.~\ref{Fig:FloquetKickQuench}(a)]
\begin{align}~\label{Eq:periodickick}
V(t) = V_{12} \; \Gamma_4 \sum^{\infty}_{r=1} \; \delta \left( t- r \; T \right),
\end{align}
where $V_{12}= \Delta \left( C_1-C_2 \right)$, $r$ is an integer and $\Gamma_4=\sigma_1 \tau_1$.

If we neglect the time dependence of $V(t)$, then the Hamiltonian $H_{\rm Stat}=H_{\rm SHI}+ \Gamma_4 V_{12}$ describes a static HOTI. Since $\{ H_{\rm SHI} , \Gamma_4 \}=0$, the term proportional to $\Delta$ acts as a mass for chiral edge states, and breaks the $C_4$ as well as time-reversal (generated by ${\mathcal T} =i\sigma_2\tau_0 K$, where $K$ is the complex conjugation) symmetries. It changes sign four times across the corners of a square lattice system. Then a generalized Jackiw-Rebbi index theorem~\cite{Rackiw-Rebbi, agarwala} ensures the existence of four zero-energy corner modes, and we realize a static second order topological insulator.

The robustness of such zero modes can be ensured from the spectral or particle-hole symmetry of $H_{\rm Stat}$, generated by the unitary operator $\Gamma_5$, as $\{H_{\rm Stat} ,\Gamma_5\}=0$, where $\Gamma_5=\sigma_2 \tau_1$. The zero-energy corner modes are eigenstates of $\Gamma_5$ with eigenvalues $+1$ and $-1$. We also identify an antiunitary operator $A=\Gamma_1 K$, such that $\{H_{\rm Stat} ,A\}=0$~\cite{roy-antiunitary}. In the following discussion on Floquet HOTI, this antiunitary operator plays an important role, about which more in a moment.

To demonstrate the possible realization of a Floquet HOTI in the presence of a periodic kick [see Fig.~\ref{Fig:FloquetKickQuench}(a) and Eq.~(\ref{Eq:periodickick})], we focus on the corresponding Floquet operator
\begin{align}~\label{Eq:Floquetoperator}
U(T) &= {\rm TO}\left( \exp \left[ -i\int_0^T \left[H_{\rm SHI} + V(t) \right] dt \right] \right)  \nonumber \\
     &= \exp(-i H_{\rm SHI} \; T) \; \exp(-i V_{xy} \Gamma_4),
\end{align}
where `$\rm TO$' stands for the time-ordered product. The Floquet operator after a single kick takes a compact form
\begin{eqnarray}~\label{Eq:Floquet}
U(T) =C_T \left[ n_0 -i n_4 \Gamma_4 \right] -i S_T \sum^3_{j=1} \left[ n_j \Gamma_j + m_{j} \Gamma_{j4} \right],
\end{eqnarray}
where $\Gamma_{jk}=[\Gamma_j, \Gamma_k]/(2i)$, $C_T= \cos (|{\bf N}({\bf k})| T)$, $S_T= \sin (|{\bf N}({\bf k})| T)$, $n_0 = \cos (V_{12})$, $n_4 = \sin(V_{12})$, and
\begin{align}
(n_j,m_j)=N_j ({\bf k}) (\cos (V_{12}), \sin (V_{12}))/|{\bf N}({\bf k})| \nonumber
\end{align}
for $j=1,2,3$. The effective Floquet Hamiltonian ($H_{\rm Flq}$) can be obtained from the relation $U(T)=\exp(-i H_{\rm Flq} T) \approx 1- i H_{\rm Flq} T + {\mathcal O} (T^2)$. In the high-frequency limit ($T \to 0$), one can neglect the higher-order terms in $T$, and comparing with Eq.~(\ref{Eq:Floquet}), we find~\cite{comment-1}
\begin{equation}~\label{Eq:FloquetHamiltonian}
H_{\rm Flq} = \sum^3_{j=1} N_j ({\bf k}) \Gamma_j + V_{12} \; \sum^3_{j=1} N_{j} ({\bf k}) \Gamma_{j4} + \frac{V_{12}}{T} \; \Gamma_4.
\end{equation}
While arriving at the final expression we assumed that $T, \Delta \to 0$, but $\Delta/T$ is finite. Notice that $H_{\rm Flq}$ looses the spectral symmetry with respect to the unitary operator $\Gamma_5$, but still satisfies $\{H_{\rm Flq}, A \}=0$. This observation ensures the spectral symmetry among Floquet quasienergy modes, and suggests possible realization of corner localized zero quasienergy modes and a Floquet HOTI.

We now anchor this anticipation by numerically diagonalizing the Floquet operator from Eq.~(\ref{Eq:Floquetoperator}), satisfying
\begin{align}~\label{Eq:quasienergy_def}
U(T) \; |\phi_n \rangle =\exp(i \mu_n T) \; |\phi_n \rangle,
\end{align}
where $|\phi_n \rangle$ is the Floquet state with quasienergy $\mu_n$, with open boundaries in both directions. In Fig.~\ref{Fig:FloquetHOTI_Summary}(d), we show the local density of states associated with the (almost) zero [${\mathcal O} (10^{-6})$] quasienergy Floquet states, which depicts a strong corner localization. Therefore, by means of a periodic kick, a Floquet HOTI can be generated from a first-order topological insulator (QSHI in this case), when the driving perturbation breaks the desired symmetries ($C_4$ and ${\mathcal T}$ here) and satisfies a specific algebraic relation ($\{ H_{\rm SHI}, \Gamma_4 \}=0$ in this case).

The topological robustness of the corner modes can be tested by computing the associated topological charge
\begin{align}
Q= \frac{1}{4} \:\: \sum_{p, q \in \{ \mu_0 \}}  \langle \phi_p |A| \phi_q \rangle,
\end{align}
measuring the overlap of a zero quasienergy state (eigenstate of $A$ with eigenvalue $+1$ or $-1$) with the states within the subspace of four zero quasienergy modes $\{ \mu_0 \}$, after acted by $A$ from left. In the HOTI phase $Q=1$ by construction, since the spectral symmetry of $H_{\rm Flq}$ generated by the antiunitary operator $A$, leaves $\{ \mu_0 \}$ invariant and $A|\phi_n \rangle$ is characterized by quasienergy $-\mu_n$. We indeed find $Q=1$ (within numerical accuracy), confirming that the zero quasienergy corner modes are stable, eigenstates of $A$, separated from bulk states with $|\mu_n|>0$, and they are topologically protected.

\emph{Floquet quadrupolar moment ($Q^{\rm Flq}_{xy}$)}. A static two-dimensional HOTI possesses a quantized quadrupolar moment $Q_{xy}=1/2$ in both crystalline~\cite{benalcazar2017, hughes-Qxy,cho_Qxy} and amorphous~\cite{agarwala} systems. We now compute the quadrupolar moment for Floquet HOTI in the following way. Notice that the Floquet modes reside within a quasienergy window $(-\omega/2,\omega/2)$, where $\omega=2\pi/T$ is the kick frequency. We work in the high frequency regime such that $\omega \gg t_{0,1}$, and compute the following quantity
\begin{equation}
n_{\rm Flq}= {\rm Re} \left[- \frac{i}{2\pi}{\rm \bf Tr} \left( \ln \left[ U^\dagger \exp\left[ i 2 \pi \sum_{\bf r} \hat{q}_{xy} ({\bf r}) \right] U \right] \right) \right],
\end{equation}
where $\hat{q}_{xy}({\bf r})= \frac{x y}{L^2} \; \hat{n}({\bf r})$ and $\hat{n}({\bf r})$ is the number operator at ${\bf r}=(x,y)$. The matrix $U$ is constructed by columnwise arranging the eigenvectors $| \phi_n \rangle$ with quasienergy $\mu_n$ [see Eq.~(\ref{Eq:quasienergy_def})], such that $-\frac{\omega}{2} < \mu_n < 0$. The Floquet quadrupolar moment (modulo 1) is then defined as
\begin{equation}
Q^{\rm Flq}_{xy}=n_{\rm Flq}-n^{\rm al}_{\rm Flq},
\end{equation}
where $n^{\rm al}_{\rm Flq}=\frac{1}{2L^2}\sum_{\bf r} (x y)$ is the value of $n_{\rm Flq}$ in the atomic limit. Inside the Floquet HOTI phase (when $|m|<2$) we find $Q^{\rm Flq}_{xy}=0.5$, whereas $Q^{\rm Flq}_{xy}=0$ for a trivial insulator (when $|m|>2$), see Fig.~\ref{Fig:FloquetQuadrupolar}(a). These features are insensitive of the system size, see Fig.~\ref{Fig:FloquetQuadrupolar}(b). A slight origin dependence of $Q^{\rm Flq}_{xy}$ inside the Floquet HOTI [see Fig.~\ref{Fig:FloquetQuadrupolar}(c)] disappears as the system approaches the thermodynamic limit $L \to \infty$, see Fig.~\ref{Fig:FloquetQuadrupolar}(d). Moreover, for trivial insulator, $Q_{xy}^{\rm Flq}$ always stays at $0$, irrespective of the choice of origin. Therefore, quantized Floquet quadrupolar moment serves as the indicator for a two-dimensional Floquet HOTI~\cite{watanabe_Qxy_problem}.

\emph{Quench dynamics}. Upon Floquet engineering a HOTI, we now investigate the \emph{survival probability} of a corner mode at time $t>0$, after \emph{suddenly} switching off the $C_4$ and ${\mathcal T}$ symmetry breaking perturbation at $t=0$ [see Fig.~\ref{Fig:FloquetKickQuench}(b)]. This process is parametrized by
\begin{align}~\label{Eq:Profile_Suddenquench}
V(t) =V_{12} \; \Gamma_4 \: \left[ 1 - \Theta(t) \right],
\end{align}
where $\Theta$ is the heaviside step function of its argument. The survival probability at time $t$ is defined as~\cite{Diptiman_book, Adutta, sacramento, tanay}
\begin{align}~\label{Eq:SurvivalProb_Def}
P_s(t)=\big| \sum_{n=1}^{4 L^2}|\langle\Psi^{\rm initial}_{\rm corner}|\Phi^{\rm final}_{n} \rangle|^2e^{-iE_nt} \big|^2,
\end{align}
where $|\Psi^{\rm initial}_{\rm corner} \rangle$ is one of the zero-energy corner modes of the initial Hamiltonian $H_{\rm Ini}=H_{\rm SHI}+ \Gamma_4 V_{12}$ for $t<0$, $|\Phi^{\rm final}_{n} \rangle$ is a wavefunction of the final Hamiltonian (after the sudden quench at $t=0$) $H_{\rm Fin}=H_{\rm SHI}$ with energy $E_n$, and $L^2$ is the total number of lattice sites in the real space. The complete, partial and no revival respectively correspond to the situations when the survival probability acquires maximum (close to unity), intermediate (but finite) and very small (close to zero) values, see Fig.~\ref{Fig:survivalprob_summary}.

The dynamics of the survival probability can be understood from the density profile of the time evolved state $|\Psi(t) \rangle = \exp(-i H_{\rm Fin} t) |\Psi^{\rm initial}_{\rm corner} \rangle$ at various instances after the quench, see Fig.~\ref{Fig:wavefunction_quench}. For concreteness, we select one of the four \emph{near} (due to finite system size) zero-energy modes, dominantly localized near the corner at $(L,L)$ [identified as $|\Psi^{\rm initial}_{\rm corner} \rangle$ in Eq.~(\ref{Eq:SurvivalProb_Def})], see Fig.~\ref{Fig:wavefunction_quench}(a). As the final Hamiltonian ($H_{\rm SHI}$) accommodates gapless one-dimensional edge modes [see Fig.~\ref{Fig:FloquetHOTI_Summary}(b)], the initial corner state predominantly diffuses along the edges of the system at $t>0$. Since the $C_4$ symmetry is restored for $t>0$, $v_x=v_y\equiv v$, where $v_i=\partial |E({\bf k})|/\partial k_i$ is the group velocity in the $i$-direction, with $i=x,y$ and $\pm E ({\bf k})$ are the eigenenergies of $H_{\rm SHI}$. The maximal value of the group velocity is $v_{\rm max} = (1+m)t^2_0/\sqrt{t^2_1 + t^2_0 (1+m)^2} \approx 1$ for $t_1=t_0=m=1$, which sets the velocity of the corner mode following the sudden quench~\cite{supplementary}.  After a time $t=L/v_{\rm max} \approx L$, the most dominant peak of the initial corner mode reaches $(1,L)$, see Fig.~\ref{Fig:wavefunction_quench}(b). The spectral density then exhibits a weaker peak at $(L,L)$. As a result, the survival probability shows an intermediate revival when $t \approx L$, see Fig.~\ref{Fig:survivalprob_summary}.

At time $t \approx 2 L$, the corner mode encounters a substantial reduction of spectral weight at $(L,L)$ and concomitantly $P_s$ shows (almost) no revival, see Fig.~\ref{Fig:survivalprob_summary}. Such behavior arises from the fact that $| \Psi (t \approx 2L) \rangle$ is (almost) orthogonal to $|\Psi^{\rm initial}_{\rm corner} \rangle$, compare Figs.~\ref{Fig:wavefunction_quench}(a) and \ref{Fig:wavefunction_quench}(c). The next partial revival occurs at $t \approx 3 L$, and a complete revival takes place at $t \approx 4 L$, when the initial state (almost) returns to itself. These features in the survival probability are impervious to the system size, group velocity ($v_{\rm max}$), obtained by tuning the hopping parameters ($t_1$ and $t_0$)~\cite{supplementary}. At later times this pattern continues to repeat itself. But, after each such cycle the amplitudes of the revival become weaker. Therefore, even after a sudden quench from a HOTI to a first-order topological insulator, the corner mode leaves its fingerprint in the survival probability in future time. By contrast when the system is quenched into a HOT insulator from a QSHI, the survival probability of the edge mode does not reveal any specific structure, possibly due to the absence of any extended gapless mode for $t>0$~\cite{supplementary}.

\emph{Discussion}. To summarize, we demonstrate a possible realization of a two-dimensional Floquet HOTI, supporting anitiunitary symmetry protected zero quasienergy corner modes [see Fig.~\ref{Fig:FloquetHOTI_Summary}(d)], by periodically kicking a QSHI (first-order topological insulator) with a discrete symmetry breaking \emph{mass} perturbation. This mechanism can be generalized to three-dimensional topological insulators and semimetals~\cite{calugaru2019}. Our proposed protocol for generating dynamic HOT phases can in principle be realized in cold atomic systems, in the presence of dynamic strain that can be generated by gluing the sample with a piezoelectric material, vibrating at high frequency~\cite{inverse-piezo}, and in acoustic~\cite{Exp-2, krishanu-acousticHOTI} and photonic~\cite{photonic-floquet} systems.

Finally, we show that the signature of the corner modes can persist for a long time after a sudden quench into a QSHI, which manifests through periodic appearances of partial and complete revival in the dynamics of the survival probability [see Figs.~\ref{Fig:survivalprob_summary} and ~\ref{Fig:wavefunction_quench}]. The predicted quench dynamics can possibly be observed in cold atomic systems~\cite{quench-RMP}, and generalizations of this scenario to higher-dimensional HOT phases are left for a future investigation. We hope that present discussion will motivate future theoretical and experimental works exploring the dynamical properties of HOT phases.

\emph{Acknowledgments}. BR is thankful to Adhip Agarwala, Takashi Oka, and Andr$\acute{\mbox{a}}$s Szab$\acute{\mbox{o}}$ for useful discussions. BR was partially supported by the Startup Grant from Lehigh University.


\begin{thebibliography}{}
\bibitem{comment-1} One can also find the same expression for $H_{\rm Flq}$ starting from Eq.~(\ref{Eq:Floquetoperator}), and subsequently using the Baker-Campbell-Hausdorff relation $\exp(x)\exp(y)=\exp(x+y+[x,y]/2 + [[x,y],y]/12 +\cdots)$.
\end{thebibliography}
\end{document}